\documentclass[aps,superscriptaddress,footinbib,twocolumn]{revtex4-1}
\usepackage{graphicx}
\usepackage{amssymb}
\usepackage{amsmath}
\usepackage{times}
\usepackage{soul}
\usepackage[outdir=./]{epstopdf}
\usepackage{xcolor}

\begin{document}

\title{Phase chimera states on non-local hyperrings}

\author{Riccardo Muolo}
\email{corresponding author: muolo.r.aa@m.titech.ac.jp}
\affiliation{Department of Systems and Control Engineering, Tokyo Institute of Technology, Tokyo 152-8552, Japan}
\affiliation{Department of Mathematics, University of Namur, B5000 Namur, Belgium}
\affiliation{naXys, Namur Institute for Complex Systems, University of Namur, B5000 Namur, Belgium}

\author{Thierry Njougouo} 
\affiliation{naXys, Namur Institute for Complex Systems, University of Namur, B5000 Namur, Belgium}
\affiliation{Faculty of Computer Science, University of Namur, B5000 Namur, Belgium}
\affiliation{Department of Electrical and Electronic Engineering, University of Buea, P.O. Box 63, Buea, Cameroon}

\author{Lucia Valentina Gambuzza} 

\affiliation{Department of Electrical, Electronics and Computer Science Engineering, University of Catania, 95125 Catania, Italy}

\author{Timoteo Carletti}
\affiliation{Department of Mathematics, University of Namur, B5000 Namur, Belgium}
\affiliation{naXys, Namur Institute for Complex Systems, University of Namur, B5000 Namur, Belgium}

\author{Mattia Frasca }
\affiliation{Department of Electrical, Electronics and Computer Science Engineering, University of Catania, 95125 Catania, Italy}
\affiliation{Istituto di Analisi dei Sistemi ed Informatica “A. Ruberti”, IASI-CNR, 00185 Roma, Italy}

\date{\today}

\begin{abstract}
Chimera states are dynamical states where regions of synchronous trajectories coexist with incoherent ones. A significant amount of research has been devoted to study chimera states in systems of identical oscillators, non-locally coupled through pairwise interactions. Nevertheless, there is an increasing evidence, also supported by available data, that complex systems are composed by multiple units experiencing many-body interactions, that can be modeled by using higher-order structures beyond the paradigm of classic pairwise networks. 
In this work we investigate whether phase chimera states appear in this framework, by focusing on a novel topology solely involving many-body, non-local and non-regular interactions, hereby named {\em non-local $d$-hyperring}, being $(d+1)$ the order of the interactions. We present the theory by using the paradigmatic Stuart-Landau oscillators as node dynamics, and show that phase chimera states emerge in a variety of structures and with different coupling functions. For comparison, we show that, when higher-order interactions are "flattened" to pairwise ones, the chimera behavior is weaker and more elusive. 
\end{abstract}

  \maketitle
  
\paragraph*{Introduction.} Chimera states are an intriguing dynamical phenomenon occurring in systems of coupled oscillators where some units oscillate synchronously, forming coherent domains that exist alongside other domains characterized by incoherent oscillations. Since their first numerical observation by Kaneko \cite{kaneko1}, the emergence of these patterns has raised the interest of many scholars in nonlinear science. In later studies, these patterns have been identified in experimental setups involving Josephson junctions \cite{cerdeira_prl},  laser systems \cite{hagerstrom2012experimental}, mechanical systems \cite{martens2013chimera}, electronic circuits \cite{vale_chimeras,gambuzza2020experimental}, nano-electromechanical oscillators \cite{matheny2019exotic}, neuroscience \cite{chimera_neuro}, unihemispheric sleep in birds, mammals, and reptiles \cite{rattenborg2000behavioral}, as well as various forms of pathological brain states \cite{majhi2019chimera}. They have been also found in numerical experiments involving systems of non-locally coupled oscillators, such as Ginzburg-Landau systems, Rössler oscillators, logistic maps \cite{kuramoto1,kuramoto2,kuramoto3,kuramoto4,kuramoto5}, and identical phase oscillators \cite{kuramoto_batt}, prior to being named ``Chimera states" in \cite{abrams2004chimera}.

Chimera states have been extensively investigated over the past years, leading to exploration along various intriguing avenues \cite{zakharova_chimeras,omel2018}. From a theoretical point of view, there has been considerable effort to characterize the different types of chimera states that may appear: from phase chimeras \cite{kuramoto_batt} to amplitude chimeras \cite{sathiyadevi2018stable}, amplitude mediated chimeras \cite{sethia2013amplitude}, chimera death states \cite{zakharova2014chimera}, traveling chimeras \cite{simo2021traveling}, and  globally clustered chimeras \cite{sheeba2010_clustered_chimera,sheeba2009globally}. Let us observe that the recently defined phase chimera~\cite{phase_abrams} differ from the ones presented in~\cite{kuramoto_batt}, despite bearing the same name. Indeed, by writing the dynamics of the $j$-th oscillator as $a_j(t)\mathrm{exp}\!\left[i(2\pi\Omega_j t +\theta_j)\right]$, Kuramoto and Battogtokh~\cite{kuramoto_batt} defined a phase chimera as a state where $a_j(t)$ is constant for all $j$ while the angles $\phi_j= 2\pi\Omega_j t +\theta_j$ can be divided into two groups, one for which $\Omega_j$ is constant and a second one for which $\Omega_j$ depends on the node index. Both behaviors can be appreciated after a sufficiently large $t$. On the other hand, Zajdela and Abrams~\cite{phase_abrams} defined a phase chimera as a state for which both $a_j(t)$ and $\Omega_j$ are constant, i.e., do not depend on the node index, and only the phases $\theta_j$ exhibit coherent and incoherent behaviors. The latter is the framework we will hereby consider.

Another avenue of research on chimera states has focused on understanding the network topologies and coupling mechanisms that can facilitate their emergence. Although chimera states have been proved to arise in $1D$-rings where every node is connected to its two neighbors (a condition named {\em local coupling}~\cite{local1,local2}) as well in the ``opposite'' case of a complete network where each node is connected to all the other ones ({\em global coupling} \cite{global1,global2}), the {\em non-local coupling} configuration proved to be particularly important for the onset of chimera states. Indeed, a large amount of literature has demonstrated that chimera states are commonly observed within such setting \cite{kuramoto_batt,omel2018}. The non-local coupling corresponds to a $1D$-ring where, beside first order neighbors, each node is also connected in a regular way to $2(k-1)\geq 2$ other ones, i.e., each node is connected to $k$ nearby nodes counted in anti-clockwise manner and $k$ in the other direction. The network is regular by construction, all the nodes having the same degree $2k$, and it has a complete group of automorphisms, i.e., it is invariant for any possible translation. Beside those regular cases, chimera states have been also found on networks with diverse non-regular topologies~\cite{nonregular,meena_str_chimera,bera2017chimera_topologies,muolo_2023_chimera,simo2021chimera, Thierry_synch}. Nonetheless, the identification of chimera states has been proven to be challenging, given their transient nature \cite{panaggio2015chimera} and strong dependence on the initial conditions \cite{chim_initial_cond}: hence, researchers have also endeavored to propose mechanisms that can lead to the emergence of robust and persistent chimeras \cite{zhang_mott_prl,bram_malb_chim}.

While there have been numerous studies focusing on chimera states within network structures, these states have received significantly less attention in systems where the units interact not only in pairs, but also in groups of three or more units. Studying chimera states in the presence of these types of interactions, known as higher-order or many-body \cite{battiston2020networks,natphys}, is particularly important. In fact, on the one hand, there is an increasing evidence that many systems, e.g., in neuroscience \cite{petri2014homological,sizemore2018cliques,Bassett,sporns_edge}, naturally exhibit higher-order interactions. On the other hand, several studies on dynamics emerging in hypergraphs and simplicial complexes have already shown a significant impact of higher-order interactions on the collective behavior of the system; the latter include, e.g., studies on synchronization and pattern formation
\cite{gambuzza2021stability,lucas2020multiorder,bick_explosive,skardal2019abrupt,skardal2020higher,leon22a,carletti2020dynamical,muologallo}, random walks \cite{carletti2020random,schaub2020random} and contagion processes \cite{iacopini2019simplicial,de2019social,stonge2021universal,Chowdhary21,JF_contagion_23}. Previous works on chimera states in higher-order structures have already lead to several interesting results \cite{matheny2019exotic,zhang2021unified,raissa_beyond_projected,ghosh_chimera_high-order,ghosh_chimera2}. {  Chimera states have been found experimentally by considering triadic interactions \cite{matheny2019exotic} and, theoretically,} while developing a general theory for the study of synchronization patterns in higher-order systems \cite{zhang2021unified,raissa_beyond_projected}; despite dealing with small hypergraphs and specific settings, these works have pointed out that chimera states may also appear in the presence of higher-order interactions. A more systematic study of the impact of higher-order interactions has been done in \cite{ghosh_chimera_high-order,ghosh_chimera2}, where phase oscillators coupled through simplicial complexes are studied; both works show that higher-order interactions, when added to paiwise ones, enhance the likelihood of observing chimera states in the  system.

In this work, we further corroborate the claim that chimera states are boosted by the presence of many-body interactions by focusing on a higher-order structure that we call \textit{non-local hyperring} and consists exclusively of higher-order interactions. This is an important difference with respect to the setup studied, for instance, in \cite{ghosh_chimera2}{ , where simplicial complexes are considered, or in \cite{bick_nonlocal1}, where a generalization of the non-local coupling is extended to higher-order interactions through a continuous limit \footnote{The same setting has further investigated in \cite{bick_nonlocal2}.}: in both settings,} the effects of pairwise and three-body interactions cannot be fully disentangled. The topology that we consider is non-local and non-regular, i.e., nodes have different hyper degree. We consider a $d$-uniform hypergraph, where all the hyperedges have the same size, $d$, and are connected two-by-two, in a periodic structure, by junctions nodes shared by every two consecutive hyperedges. In this way, we have a non-regular extension of the notion of non-local rings to higher-order structures. Since non-local hyperrings only include higher-order interactions, the effect of the latter on the emergence of chimera states can be better identified. Let us observe that, despite the lack of regular topology, the studied systems always admit a global synchronous solution, that thus ``competes'' with the possible chimera state. For the sake of definiteveness, we numerically investigate the behavior of non-local hyperrings when the nodal dynamics is given by the Stuart-Landau oscillator, {  which has been widely used as a paradigmatic model of oscillatory dynamics to study chimera states \cite{zakharova_chimeras}, given that this system is the normal form of the Hopf-Andronov bifurcation and thus it presents the general features of any limit-cycle oscillator close to such bifurcation \cite{Nakao_SL}.} We then compare the results with the ones observed on a pairwise network, obtained by projecting the higher-order structure on a pairwise network: namely, we consider that every node has a non-weighted pairwise connection with all the nodes part of the same hyperedge, i.e., they form a clique. The higher-order setting exhibits a considerable enhancement of chimera patterns, meaning that they are present for a wider range of coupling strengths and they have a longer life span. Finally, we show that not only phase chimera states, but also other interesting dynamical patterns, such as amplitude chimeras \cite{zakharova2016amplitude}, can be observed in the novel setting. 

\paragraph*{Stuart-Landau oscillators
%with nonlinear coupling
on non-local hyperrings.} We consider a system made of $n$ interacting Stuart-Landau units{ , a paradigmatic model of oscillatory dynamics}. In the absence of any interaction, each unit $j$ ($j=1,\ldots,n$) of the system is described by the following equations
\begin{equation}
   \begin{cases}
 \dot{x}_j = \lambda  x_j - \omega y_j  -\left(x_{j}^2  + y_{j}^2 \right)x_{j}=f(x_j,y_j), \\
 \dot{y}_j = \omega x_j + \lambda y_j - \left(x_{j}^2  + y_{j}^2 \right)y_{j}=g(x_j,y_j)\, , \end{cases}
 \label{eq:SL}
\end{equation}
where $\lambda$ is a bifurcation parameter controlling the onset of a limit cycle of amplitude $\sqrt{\lambda}$, for $\lambda>0$, and $\omega$ is the natural frequency of the oscillator. Let us observe that we assume those parameters to be same for all nodes to ensure the existence of a global synchronous solution.

Here, as in \cite{neuhauser2020multibody,gambuzza2021stability}, we study nonlinear many-body coupling functions that cannot be decomposed into a combination of weighted pairwise interactions \footnote{Let us observe that the nonlinearity of the coupling functions may not be enough to deal with proper higher-order interactions. In fact, as shown in \cite{muologallo}, if the nonlinear functions $h$ are the sum of terms $h(x_\ell)-h(x_j)$, they can still be decomposed onto pairwise ones. In what follows, we chose the coupling functions such that the latter case does not occur.}.
To model the higher-order interactions we use hyperedges, whose structure can be encoded by using adjacency tensors, that are a generalization of the adjacency matrix for networks \cite{battiston2020networks}. We adopt the convention that a hyperedge involving $(d+1)$ nodes (and, thus, encoding a $(d+1)$-body interaction) is called a $d$-hyperedge. Such notation is more common in the literature dealing with simplicial complexes \cite{bianconi2021higher}, while, in the context of hypergraphs, often a $d$-hyperedge encodes a $d$-body interaction \cite{battiston2020networks}. For example,  $A^{(3)}=\{A_{i,j,k,l}^{(3)}\}$ is the $3$-rd order adjacency tensor, encoding the $4$-body interactions, with $A_{i,j,k,l}^{(3)}=1$ if units $i,j,k,l$ have a group interaction (namely, nodes $i,j,k,l$ are part of the same $3$-hyperedge), and $0$ otherwise. Using these tensors, the generalized $d$-degree (or hyperdegree), $k_{j}^{(d)}$, representing the number of $d$-hyperedges of which node $j$ is part, can be computed as
\begin{equation}\label{eq:ddeg}
k_{j}^{(d)}=\frac{1}{d!}\sum\limits_{j_1,..,j_d=1}^N A_{jj_1\dots j_d}^{(d)}.
\end{equation}

\noindent Finally, we assume that the coupling terms act on both the dynamical variables describing the Stuart-Landau oscillator \eqref{eq:SL} and involve only the first component of the state vector of the oscillator, as in \cite{gambuzza2020experimental};  {let us observe that this working assumption is not restrictive and other couplings can also be considered}. Taking into account all these considerations, the model of Stuart-Landau equations coupled via a generic $(d+1)$-body interactions reads:
\begin{equation}
\begin{cases}
\displaystyle \dot{x}_j =f(x_j,y_j) + \epsilon \sum_{j_1,...,j_d=1}^n A_{j,j_1,...,j_d}^{(d)}\Big(h^{(d)}(x_{j_1},...,x_{j_d}) \\~~~~~~~~~~~~~~~~~~~~~~~~~~~~~~~~~~~~~~~~~~~~~~~~~~~~~~~~~~~~~~ - h^{(d)}(x_j,...,x_j) \Big),\\ \\
\displaystyle \dot{y}_j =g(x_j,y_j)+ \epsilon \sum_{j_1,...,j_d=1}^n A_{j,j_1,...,j_d}^{(d)}\Big(h^{(d)}(x_{j_1},...,x_{j_d})\\~~~~~~~~~~~~~~~~~~~~~~~~~~~~~~~~~~~~~~~~~~~~~~~~~~~~~~~~~~~~~~ - h^{(d)}(x_j,...,x_j) \Big),
\end{cases}
\label{eq:SL_HO4}
\end{equation}
where $\epsilon >0$ is the coupling strength and we assumed that the coupling is diffusive-like~\cite{gambuzza2021stability}. The model encompasses only multi-body interactions (specifically, $(d+1)$-body interactions encoded in the $d$-th order adjacency tensor $A^{(d)}$), as we aim to analyze a pure many-body framework. 

Let us observe that higher-order interactions are often considered together with pairwise interactions \cite{ghosh_chimera_high-order,ghosh_chimera2}; however, in this setting it is not clear how to distinguish the different types of contributions. For this reason, we decided to consider hypergraphs instead of simplicial complexes, as the latter structures contain interactions of any order, from the highest to the lowest (i.e., order one, which is the pairwise case). Moreover, to the best of our knowledge, structures representing the higher-order counterpart of non-local rings have not yet been considered in previous works on chimera states. Our goal is to fill this gap by considering a ring encompassing $m$ hyperedges of order $d$. These hyperedges are placed on a circle and labeled in counterclockwise order as $1,2,\ldots,m$. In the simplest case, the $m$ hyperedges share with each other only a single node. Furthermore, each shared node may belong at most to two different hyperedges, meaning that the generic hyperedge $j$ has two shared nodes, one with the hyperedge $(j-1)$ and one with the hyperedge $(j+1)$. The remaining $d-2$ nodes are only part of the hyperedge $j$. Let us observe that more general structures could be handled as well.

\begin{figure}[htp!] 
\centering
\includegraphics[width=0.5\textwidth]{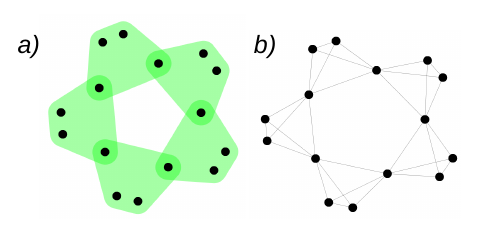}
\caption{\textbf{Non-local hyperring vs. clique-projected network.} a) A non-local $3$-hyperring with $m=5$ hyperedges {  and $15$ nodes.} b) Corresponding clique-projected network obtained by transforming each hyperedge into a clique.}
\label{fig::coupl}
\end{figure}

In the following, we will focus on the case of $4$-body interactions, but similar results have been obtained for $3$-body, as well as for $5$-body and $6$-body interactions (see Supplementary Material (SM)). In our model, the coupling functions have been chosen to be cubic and diffusive-like, as previously done in \cite{gambuzza2021stability}, namely 
\begin{equation}
 \label{eqh3}
h^{(3)}(x_{j_1},x_{j_2},x_{j_3})=x_{j_1}x_{j_2}x_{j_3}.  % \text{ and } h^{(3)}(x_i,x_i,x_i)=x_i^3\, .
\end{equation}
In the SM, we illustrate other choices, i.e., diffusive-like functions that are not cubic. In this specific case, an example of non-local ring with $m=5$ hyperedges of size $4$ is shown in panel a) of Fig.~\ref{fig::coupl}. The structure is non-regular, as some nodes are part of two hyperedges, i.e., they have hyperdegree $2$ (see Eq. \eqref{eq:ddeg}), while others interact only within a single hyperedge, and thus exhibit hyperdegree $1$. In general, for a non-local ring with $d$-body interactions and $n$ nodes, the nodes that are shared by two hyperedges, i.e., nodes $\{1,1+(d-1),...,n-d-(d-3), n-(d-2)\}$ have generalized $d$-th degree \footnote{Let us point out that labeling of the nodes is arbitrary.}, equal to $2$, while all the other nodes have generalized $d$-th degree equal to $1$.  In the following, we will be interested in comparing the dynamical behavior of SL oscillators coupled via higher-order structures with the one resulting from a pairwise one. To have a reliable comparison, we decide to work with the network obtained by projecting each hyperedge in a clique, as shown in panel b) of Fig.~\ref{fig::coupl}, to obtain the so called clique-projected network (cpn). Namely, each couple of nodes in the hyperedge is connected with an unweighted pairwise link. In this way, we end up with $d$-cliques connected through junction nodes, that are part of two consecutive cliques, exactly as the hyperedges are disposed in the hyperring. {  Let us point out that this is not the only possible projection onto a network, as one could also consider projection based on 1-cell complexes, which are
topological objects obtained by gluing together cells in different ways. In this sense, they can be viewed as generalizations of simplicial complexes~\cite{bianconi2021higher} that have recently found applications in several field, from machine learning \cite{hajij2020cell} to synchronization dynamics \cite{carletti2022global}. In the SM we have analysed a different projection, specifically, mapping any $d$-hyperedge as a regular polygon having $(d+1)$ vertices.} When the topology is pairwise, the coupling functions take only one variable, we thus define $h^{cpn}(x_j)=h^{(3)}(x_{j_1},x_{j_2},x_{j_3})|_{x_{j_1}=x_{j_2}=x_{j_3}=x_j}$. In our example of $4$-body interactions, the coupling functions~\eqref{eqh3} can be replaced on the clique-projected networks (cpn) by
\begin{equation*} 
h^{cpn}(x_\ell)-h^{cpn}(x_j)=x_\ell^3-x_j^3\, .
\end{equation*}

{  To make the difference between the pairwise and higher-order coupling more visible, in the SM, we explicitly write the equations for a node shared between two hyperedges (junction node) and a node belonging to only one hyperedge (non-junction node).}

\begin{figure*}[ht!]
\centering
\includegraphics[width=0.9\textwidth]{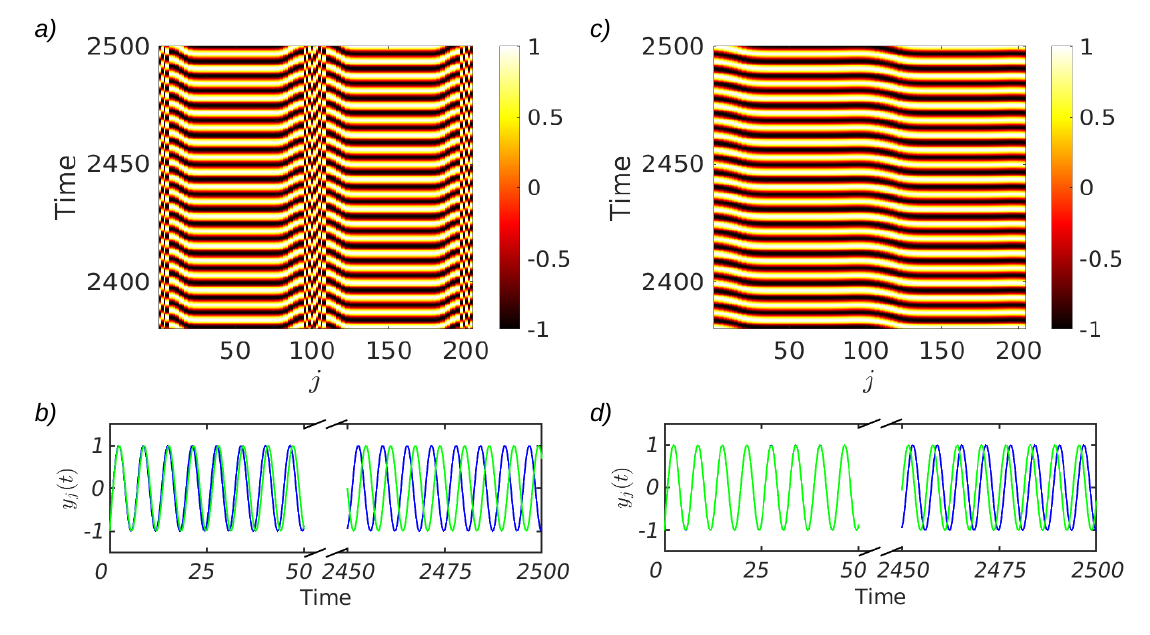}
\caption{\textbf{Non-local hyperring vs. clique-projected network.} Left panels show spatiotemporal patterns a) and time series b) for variables $y_j(t)$ of the SL oscillators coupled with a non-local $3$-hyperrings with $n=204$ nodes and $m=68$ hyperedges; the emerging dynamical behavior is a phase chimera state with two heads, i.e., there are regions of regular behavior, separated by two regions of decoherence. Right panels c) and d) show the analogous quantities on the clique-projected network. Panels b) and d) show the time series for nodes 50 (blue) and 101 (green). For panels a) and c), on the horizontal axis we set the node index $j$ while in the vertical one, the time. The coupling strength is fixed at $\epsilon=0.01$ and model parameters are $\lambda=1$ and $\omega=1$.}
\label{fig:higher_vs_pair}
\end{figure*}
\begin{figure*}[ht!]
\centering
\includegraphics[width=0.9\textwidth]{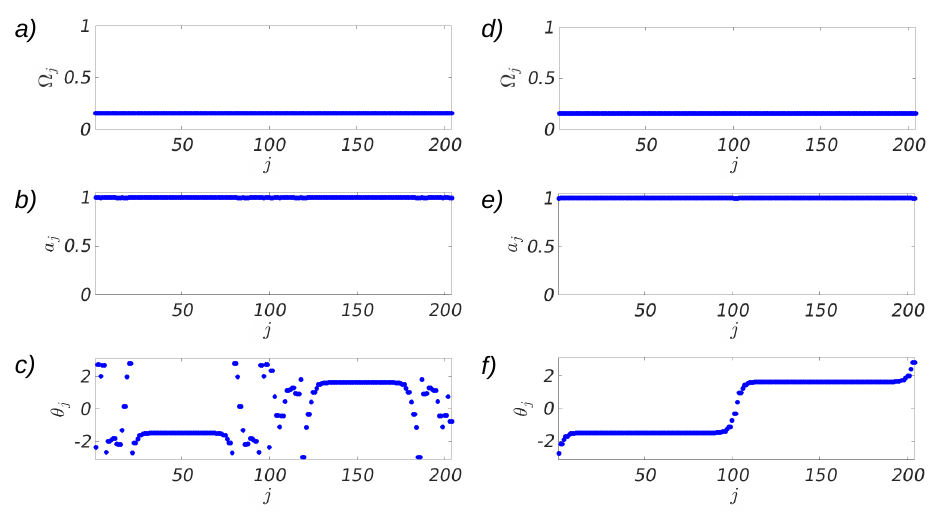}
\caption{\textbf{Dynamical quantities associated to non-local hyperring and clique-projected network.} Left panels show frequency a), amplitude b) and phase c) of the Stuart-Landau oscillators computed by using the Fast Fourier Transform (FFT) on the complex variables $z_j(t)=a_j(t)\mathrm{exp}\left[i(2\pi \Omega_j t + \theta_j)\right]$, as function of the node index, $j$, at $2500$ time units. The left panels refer to quantities whose evolution is determined by a non-local $3$-hyperring with $n=204$ nodes and $m=68$ hyperedges. The right panels show the analogous quantities for Stuart-Landau oscillators coupled via the clique-projected network. The coupling strength is $\epsilon=0.01$ and the model parameters are $\lambda=1$ and $\omega=1$.}
\label{fig:higher_vs_pair2}
\end{figure*}

\paragraph*{Higher-order vs. pairwise interactions.} We now proceed to the analysis of  chimera states observed in the non-local hyperring, while also comparing them to the pairwise case. In all the simulations, we have considered hyperrings and networks of $n=204$ nodes, hence $m=68$ hyperedges of size $4$ (i.e., $3$-hyperedges), and we have set initial conditions to $(x_i=+1, y_i=-1)$ for nodes indexes $j \leq n/2$ and $(x_j=-1, y_j=+1)$ for $j > n/2$, as in \cite{zakharova_chimeras}. 
\\\\

\noindent In Fig.~\ref{fig:higher_vs_pair}, we compare the dynamics emerging by using higher-order and pairwise interactions. In panels a) and b), we show the behavior of the system of SL oscillators coupled with a non-local $3$-hyperring, while panels c) and d) depict the same system of oscillators, but now coupled via the clique-projected network (pairwise interactions). In panel a), we can appreciate a phase chimera behavior, 
whereas in panel c) there are no domains with incoherent behaviors. 
Panels b) and d) show the time series for short and long times for nodes $j=50$ (blue) and $j=101$ (green). First of all, we observe that in both cases the orbits have the same amplitude and the same period, hence same frequency. However, in the case of the higher-order coupling, nodes initially synchronized are eventually found in phase opposition (see blue and green curves on panel b)); on the other hand, in the clique-projected network, nodes develop a small phase lag (see blue and green curves on panel d)). We are thus facing phase chimeras~\cite{phase_abrams} in the hyperring, but not on the clique-projected network. To better analyze the obtained pattern, we cast the SL variables into a single complex number, one for each node, $z_j(t):=x_j(t)+i y_j(t)$, and we then rewrite the latter as  $z_j(t)=a_j(t)\mathrm{exp}\left[i(2\pi \Omega_j t + \theta_j)\right]$, where $a_j$ is the signal amplitude, $\Omega_j$ the frequency and $\theta_j$ the constant phase.

By using the Fast Fourier Transform and a sliding window setting, we numerically compute these quantities for all nodes after a sufficiently long transient interval. The results reported in Fig.~\ref{fig:higher_vs_pair2} support the claim that we are dealing with phase chimeras according to~\cite{phase_abrams}: indeed, the amplitudes, $a_j$, and the frequencies, $\Omega_j$, are almost constant, i.e., their values do not depend on the node indexes, while the constant phases, $\theta_j$, are node-dependent. In the case of higher-order coupling (panel c)), the phases can be divided into two regions: a first one, where the phase value does not depend on the node index, and a second one, where the opposite holds true. Moreover, in this second region the phases are scattered in $[-\pi,\pi)$. The clique-projected network has a similar behavior: amplitudes and frequencies have constant values, while the phases exhibit  {smooth transitions between two ``flat'' regions, where there is no node dependence of the phase $\theta_j$}. The latter difference, which can be quantified through the \textit{normalized total phase variation} (see SM), is indeed the key: while for the system with higher-order coupling we can see a clear chimera behavior, the aforementioned smooth variation cannot be considered a chimera state, as there is no coexistence between coherent and incoherent states. {  In the SM, we show that such difference between a higher-order topology and a pairwise one persists also when considering different pairwise projections, such as the aforementioned boundary of the cell complex. } Although a full correspondence between the two cases cannot be established, as we are dealing with interactions that are intrinsically distinct, the difference is striking and it is consistently observed when the coupling strength is varied. Moreover, it also persists for different hyperrings, such as the $2$-, $4$- and $5$-hyperrings, as shown in the SM numerically and by mean of the normalized total phase variation.

\paragraph*{Conclusions.} In this work, we have studied the effect of pure many-body interactions on chimera states, showing that the presence of a higher-order topology enhances the emergence of phase chimeras. We have introduced a non-local non-regular (pure) higher-order topology, which we called non-local hyperring, and observed that the comparison between the dynamics on the latter and on the clique-projected network shows a remarkable difference in the behaviors and allows to draw a conclusion that is consistent with the existing literature, namely that higher-order interactions facilitate the emergence of chimera states in systems of coupled oscillators. Moreover, non-local hyperrings allow for the emergence of other phenomena, as we show in the SM, such as a hybrid state of amplitude chimeras and chimera death, and pure chimera death \cite{zakharova2014chimera}. We believe that the emergence of such type of chimera states, widely studied for pairwise interactions, but until now never observed with higher-order ones, makes our framework promising and will inspire future studies in this direction.

\paragraph*{Acknowledgements} 
The authors are grateful to Giovanni Reina, Elio Tuci, Hilda Cerdeira and Hiroya Nakao for useful discussions and feedback. R.M. acknowledges JSPS KAKENHI JP22K11919, JP22H00516, and JST CREST JP-MJCR1913 for financial support. During part of this work, R.M. was supported by a FRIA-FNRS Fellowship, funded by the Walloon Region, Grant FC 33443. T.N. expresses its gratitude to the University of Namur for financial support.

\onecolumngrid

\vspace{1cm}

\hrule

\setcounter{equation}{0}
\renewcommand{\theequation}{SM\arabic{equation}}
\setcounter{figure}{0}
\renewcommand{\thefigure}{SM\arabic{figure}}
\setcounter{section}{0}
\renewcommand{\thesection}{SM\arabic{section}}

\section{Normalized total phase variation}\label{sec:appA}

To describe quantitatively the phenomenon of chimera states is a difficult task. In most of the literature on the subject, the presence or not of a chimera state is depicted via visualization of the patterns, and no measure is given to quantify the phenomenon. Nonetheless, there are some exceptions: for example, Gopal et al. \cite{gopal2014} proposed an empirical measure which quantifies the coherence of the state. Such measure is able to determine the region where the system is not fully coherent, but such states are not always chimera ones. Moreover, it cannot distinguish between a phase chimera and other kinds of chimeras. For this reason, we put to use a concept of measure theory, often used in image processing \cite{rudin1990}: that of \textit{total variation} of a function \cite{totalvariation}. Such measure accounts for how much a certain function $f$ varies on a given interval $I$ and it is defined as follows 
\begin{displaymath}
    \mathcal{V}_I(f)=\sup_P \sum_{j=0}^{n-1} |f(x_{j+1})-f(x_j) | ,
\end{displaymath} where $P$ is a partition of $I$ realized with $n+1$ distinct elements of $I$, $x_0<x_1<\dots <x_{n}$. For a more rigorous definition, the reader may consult the cited references. \\
In our case, if applied to the phase, it can tell us the total variation of the phase in the system of oscillators. Our function $\theta$ is discrete (there are $n$ oscillators) and periodic ($\theta_{n+1}=\theta_1$), hence we can define its total variation as \begin{equation}
    \mathcal{V}(\theta)=\max \sum_{j=1}^{n}| (\theta_{j+1}-\theta_j)|.
\end{equation} 

Because we deal with hypergraphs and networks of different nodes, and the difference between two phases is at most $\pi$, we hearby deal with the \textit{normalized total phase variation}, namely,\begin{equation}
    V(\theta)=\frac{1}{\pi n}\mathcal{V}(\theta).
\end{equation} 
We will denote $V_\theta^{hr}$ and $V_\theta^{cpn}$ for systems of non-local hyperrings and clique-projected networks, respectively. \\
For the case of Figs. 2 and 3 of the Main Text, we have that $V_\theta^{hr}=0.1295$ and $V_\theta^{cpn}=0.0201$.

 {  \section{Explicit form of the equations for $3$-hyperrings}\label{subsec:appAbb}

Eq. (3) of the Main Text gives the dynamics of the system under study for a generic $d$ order interaction. Let us now write the equations for the system on a $3$-hyperring (i.e., $4$-body interactions):

\begin{equation}
\begin{cases}
\displaystyle \dot{x}_j =f(x_j,y_j) + \epsilon \sum_{j_1,j_2,j_3}^n A_{j,j_1,j_2,j_3}^{(3)}\Big(x_{j_1}x_{j_2}x_{j_3} - x_j^3\Big),\\ \\
\displaystyle \dot{y}_j =g(x_j,y_j)+ \epsilon \sum_{j_1,j_2,j_3}^n A_{j,j_1,j_2,j_3}^{(3)}\Big(x_{j_1}x_{j_2}x_{j_3} - x_j^3 \Big). 
\end{cases}
\label{eq:SM_eq}
\end{equation}

The analogous system on the project-clique network is the following:

\begin{equation}
\begin{cases}
\displaystyle \dot{x}_j =f(x_j,y_j) + \epsilon \sum_{j,l}^n A_{j,l}\Big(x_{l}^3 - x_j^3 \Big),\\ \\
\displaystyle \dot{y}_j =g(x_j,y_j)+ \epsilon \sum_{j,l}^n A_{j,l}\Big(x_{l}^3 - x_j^3 \Big). 
\end{cases}
\label{eq:SM_eq2}
\end{equation}

On the hyperring, there are some nodes which are shared between two hyperedges (junction nodes), while the rest of the nodes belong each to one hyperedge (non-junction nodes). The same happens for the clique-projected network, where junction nodes are shared between two cliques, while non-junction ones belong each to one clique. In both cases, the coupling functions take a different form, depending on type of node. In Fig. SM1, we display a $3$-hyperring and its clique-projected network with the explicit form of Eqs. \eqref{eq:SM_eq} and \eqref{eq:SM_eq2} for a junction and a non-junction node.}

\section{Effects of the number of nodes}\label{subsec:appAb}

We discuss what happens to the chimera states when the number of nodes is changed. We have found that the qualitative behavior is conserved, meaning that there are no or weak phase chimeras on the clique-projected networks, while they appear on non-local hyperrings. However, what changes with the number of nodes is the time span of the chimera patterns. In fact, chimera states are known to be transient \cite{zakharova_chimeras}, except for some special cases induced by specific network topologies, such as modular \cite{bram_malb_chim} and non-normal \cite{muolo_2023_chimera} networks. Our case makes no exception, and the observed phase chimeras fade away into incoherence. The lower the number of nodes, the shorter the lifetime of the chimera patterns. In Fig.~SM2, we show the patterns for non-local $3$-hyperrings of $408$, $102$, $48$ and $24$ oscillators, respectively in the same time range of Fig. 2 of the Main Text  (where the number of oscillators is $204$). The latter system still exhibits phase chimeras for $408$, $102$ and $48$ nodes (panels a-l), but we can observe that the two zones of coherence are decreasing with the size of the hyperrings; on the other hand, when the number of oscillators is $24$, we only see incoherence, because the chimera state has faded away (panels m-p). Such qualitative observation is validated by the frequencies, amplitudes and phases computed with the FFT (see analysis in the Main Text), and by the normalized total phase variation, which is $V_\theta^{hr}=0.0648$, $0.2584$, $0.4016$ and $0.4312$, for $408$, $102$, $48$ and $24$ nodes, respectively.
%%%%%%%%%%%%%%%%%%%%%%%%%%%%%%%%%%%%%%%%%%%%%%%%%%%%%%%%%%%%%%%%%%%%%%%%%%%%%%%%%%%%%%%%%%%%%%%%%%%%%%%%%%%%%%%%%%%%%%%%%%%%%%%%%%%%%%%%%%

\section{Lower and higher many-body configurations}\label{sec:appB}

We can further develop our analysis by studying non-local hyperrings of different orders and make the comparison with the corresponding clique-projected networks (cpn). We always used odd couplings, i.e., cubic, quintic, etc., to preserve the sign of the cpn coupling functions. We chose the coupling functions as in \cite{gambuzza2021stability}.

\subsection{$5$-body interactions}\label{subsec:appBa}

Let us now proceed with the case of a $4$-hyperring, shown in Fig. \ref{fig:5hyperrring}. The equations read

\begin{equation}
\begin{cases}
\displaystyle \dot{x}_j = f(x_j,y_j) + \varepsilon \sum_{j_1,j_2,j_3,j_4=1}^n A_{j,j_1,j_2,j_3,j_4}^{(4)} \left(x_{j_{1}}^2x_{j_{2}}x_{j_{3}}x_{j_{4}} - x_{j}^5 \right), \\
\displaystyle \dot{y}_j = g(x_j,y_j) +  \varepsilon \sum_{j_1,j_2,j_3,j_4=1}^n A_{j,j_1,j_2,j_3,j_4}^{(4)} \left(x_{j_{1}}^2x_{j_{2}}x_{j_{3}}x_{j_{4}} - x_{j}^5 \right).
\end{cases}
 \label{eq:SL_HO5}
\end{equation}
In Fig. \ref{fig:5body}, we show that on the hyperring chimera states can be found for certain values of the coupling strength, while the system on the clique-project network does not exhibit a region of decoherence. The latter observation is validated by the frequencies, amplitudes and phases computed with the FFT, and by the normalized total phase variation, which is $V_\theta^{hr}=0.0744$ and $V_\theta^{cpn}=0.0220$ for the $4$-hyperring and the cpn, respectively.

%%%%%%%%%%%%%%%%%%%%%%%%%%%%%%%%%%%%%%%%%%%%%%%%%%%%%%%%%%%%%%%%%%%%%%%%%%%%%%%%%%%%%%%%%%%%%%%%%%%%%%%%%%%%%%%%%%%%%%%%%%%%%%%%%%%%%%%%%%

\subsection{$6$-body interactions}\label{subsec:appBb}

For the $6$-body interactions, i.e., $5$-hyperrings, we have chosen a quintic coupling. The dynamics is thus described by the following system

\begin{equation}
\begin{cases}
\displaystyle \dot{x}_j = f(x_j,y_j) + \varepsilon \sum_{j_1,j_2,j_3,j_4,j_5=1}^n A_{j,j_1,j_2,j_3,j_4,j_5}^{(5)} \left(x_{j_{1}}x_{j_{2}}x_{j_{3}}x_{j_{4}}x_{j_{5}} - x_{j}^5 \right) \\
\displaystyle \dot{y}_j = g(x_j,y_j) +  \varepsilon \sum_{j_1,j_2,j_3,j_4,j_5=1}^n A_{j,j_1,j_2,j_3,j_4,j_5}^{(5)} \left(x_{j_{1}}x_{j_{2}}x_{j_{3}}x_{j_{4}}x_{j_{5}} - x_{j}^5 \right)
\end{cases}
 \label{eq:SL_HO6}
\end{equation}

In Fig. \ref{fig:6body}, we show that the behavior is similar to the $4$- and $5$-body case and chimera states are found on the higher-order topology, while the pairwise setting yields a coherent behavior. The latter observation is validated by the frequencies, amplitudes and phases computed with the FFT, and by the normalized total phase variation, which is $V_\theta^{hr}=0.1293$ and $V_\theta^{cpn}=0.0223$ for the $5$-hyperring and the cpn, respectively.

\subsection{$3$-body interactions}\label{subsec:appBc}

For the $3$-body case, i.e., $2$-hyperring, we chose cubic coupling functions, as in \cite{gambuzza2021stability}. The equations are

\begin{equation}
\begin{cases}
\displaystyle \dot{x}_j = f(x_j,y_j) + \varepsilon \sum_{j_1,j_2=1}^n A_{j,j_1,j_2}^{(2)} \left(x_{j_{1}}^2x_{j_{2}} - x_{j}^3 \right), \\
\displaystyle \dot{y}_i = g(x_j,y_j) +  \varepsilon \sum_{j_1,j_2=1}^n A_{j,j_1,j_2}^{(2)} \left(x_{j_{1}}^2x_{j_{2}} - x_{j}^3 \right).
\end{cases}
 \label{eq:SL_HO3}
\end{equation}
In Fig. \ref{fig:3body}, we show the results for the dynamics on a $2$-hyperring and its corresponding clique-projected network. The chimera behavior is observed on the higher-order topology, while, again, it fades away when the interactions are pairwise. The latter observation is validated by the frequencies, amplitudes and phases computed with the FFT, and by the normalized total phase variation, which is $V_\theta^{hr}=0.0867$ and $V_\theta^{cpn}=0.0194$ for the $2$-hyperring and the cpn, respectively.

{  \section{An alternative projection from a higher-order topology to a pairwise one}

As we anticipated in the Main Text, the clique-projected network is not the only possible projection from the hyperring to the network. Another interesting case that can be studied is inspired from cell complexes. Let us consider a $d$-hyperedge to be a $2$-cell of a cell complex (i.e., made of only nodes, links and faces, i.e., filled polygons), the projection will thus consist into associating to each face its boundary, i.e., the polygon or equivalently a $d+1$-cycle. Considering a $d$-hyperring composed of $m$ $d$-hyperedges, we would obtain $m$ adjacent $d+1$-cycles, which would share at most one node with two other ``consecutive'' cycles. Such a topology is shown in Fig. \ref{fig:necklace}, where we depict the projections for the case of $3$- and $4$-hyperrings. Let us observe that for a $2$-hyperring the clique-projected network and the latter projection coincide. Lastly, let us point out that, while the clique-projected network has a unique representation, this alternative projection does not. For example, in the case of the $3$-hyperring one could attach the cycles by a different vertex (looking at Fig. \ref{fig:necklace}a), one would have adjacent ``diamonds'' instead of squares and the cycle at the center would double its length).
We have performed the same analysis of the cpn on this alternative pairwise topology for the projections of $3$-, $4$- and $5$-hyperrings and we have found that, as for the case of cpn, there are no phase chimera states, as shown in Fig. \ref{fig:necklace2}. The latter observation is validated by the frequencies, amplitudes and phases computed with the FFT, and by the normalized total phase variation, which is $V_\theta^{p3}=0.0190$, $V_\theta^{p4}=0.0204$ and $V_\theta^{p5}= 0.0223$ for the projections of $3$-, $4$- and $5$-hyperrings, respectively. We have also tested the same setting on different representations of the projection and observed the same behavior. This additional result further strengthens our claim about the importance of higher-order interactions in the observation of chimera patterns. 
}

\section{Hybrid amplitude chimeras and chimera death}\label{sec:appC}

The model we have studied, i.e., coupled Stuart-Landau oscillators, is known in the literature to exhibit other kinds of chimera patterns, such amplitude chimera and chimera death \cite{zakharova2014chimera}. In a nutshell, the latter arises when, after a transient, the oscillations die and the system exhibits stationary patterns, while the former consists in oscillations that are in phase but with different amplitudes \cite{zakharova_chimeras}. Indeed, also in our framework of non-local hyperrings, we found both kinds of patterns, which are shown in Figs. \ref{fig:ampli} (amplitude chimera) and \ref{fig:death} (chimera death). Regarding the phenomenology of the amplitude chimeras, we can observe that they are not pure amplitude chimeras, but rather a hybrid state where some nodes are oscillating, while others are stationary. As they are not pure amplitude chimeras, we called them hybrid amplitude chimeras.
Both phenomena, hybrid amplitude chimeras and chimera death, are not exclusive of the non-local hyperring framework, but we also found them on the clique-projected networks. We have not conducted a thorough comparative study between the two settings as we did for phase chimeras, as this goes beyond the scope of this work. However, we thought that it would be an interesting result to mention, in order to stimulate future studies in this direction.

\newpage
%%%%%%%%%%%%%%%%%%%%%%%%%%%%%%%%%%%%%%%%%%%%%%%%%% Figures

%%%%%%%%%%%%%%%%%%%%%%%%%%%%%%% coupling functions

\begin{figure}[htp!]
\centering
\includegraphics[scale=0.2]{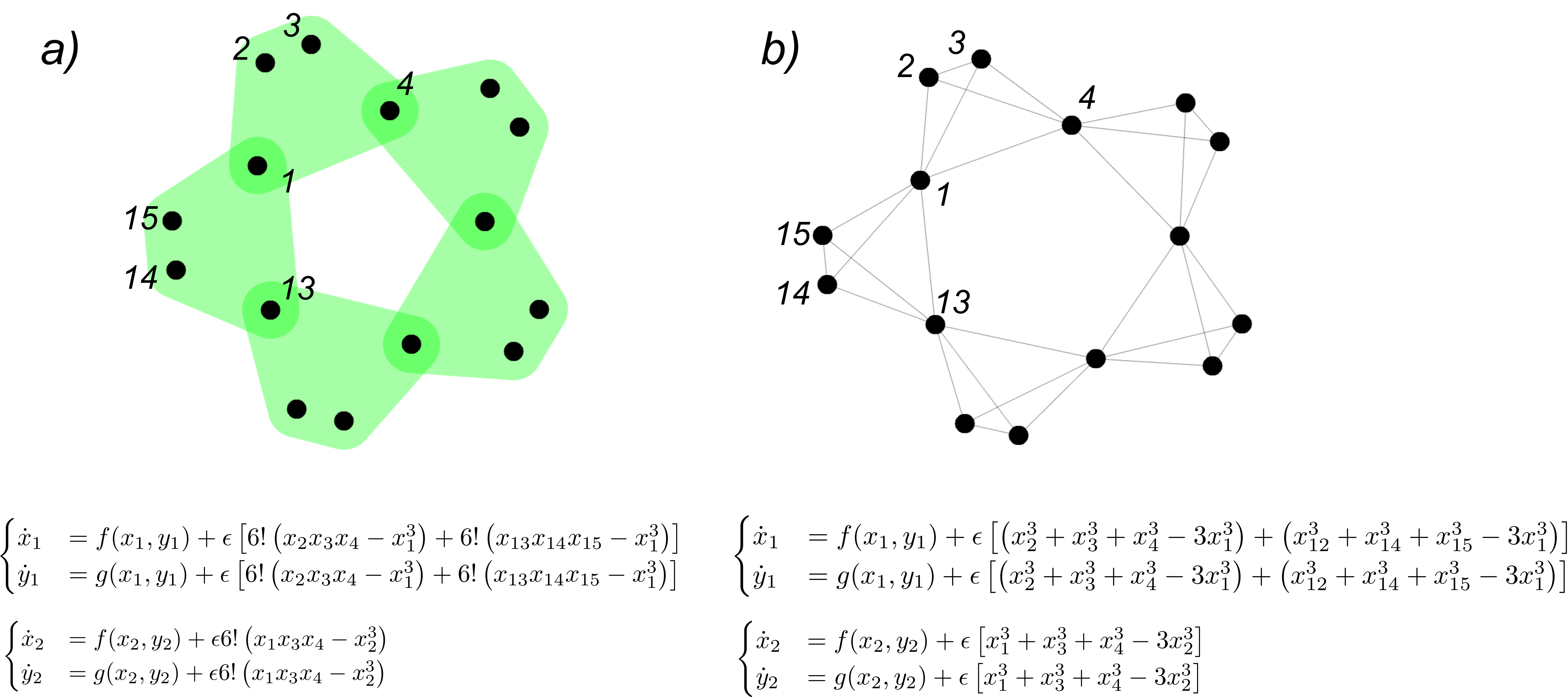}\label{figSM1}
\caption{{ \textbf{Equations for junction vs. non-junction nodes} A non-local $3$-hyperring with $m=5$ hyperedges and $15$ nodes. b) Corresponding clique-projected network obtained by transforming each hyperedge into a clique. The nodes in two hyperedges and cliques are labeled, so that the explicit form of the coupling can be displayed for a junction and a non-junction node, nodes $1$ and $2$ respectively, at the bottom of the Figure. Note that the $6!$ comes from the permutations of the indices in the coupling function, the hypergraphs being symmetric.}}
\end{figure}

%%%%%%%%%%%%%%%%%%%%%%%%%%%%%%% 4-body less/more nodes

\begin{figure}[htp!]
\centering
\includegraphics[scale=1]{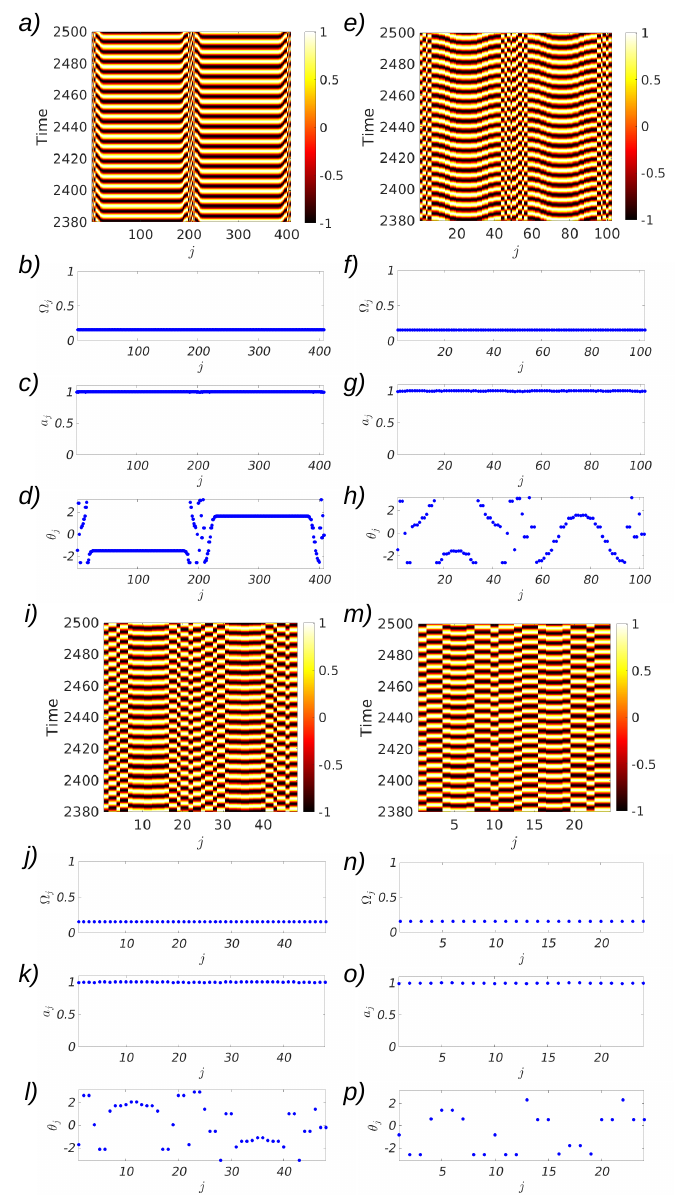}\label{aa:aaa}
\caption{\textbf{Non-local $3$-hyperring: persistence of chimera patterns vs. number of nodes} Spatiotemporal patterns for $y(t)$ and frequencies, amplitudes and phases obtained through the FFT at $2500$ time units: we observe that a chimera state with two heads is obtained with $408$ oscillators (panels a-d), and persists for $204$ (Figs. 2 and 3 of the main text), $102$ nodes (panels e-h) and even for $48$ nodes (panel i-l), but the coherent regions are becoming smaller. On the other hand, the case with $24$ oscillators (panel m-p) exhibits incoherence, because the chosen time frame is longer than the life of the chimera state, which has decayed. $j$ is the node index. For all cases, the coupling strength is $\epsilon=0.01$ and the model parameters are $\lambda=1$ and $\omega=1$.}
\end{figure}

%%%%%%%%%%%%%% 5-body topology
\begin{figure}[htp!]
\centering
\includegraphics[scale=1]{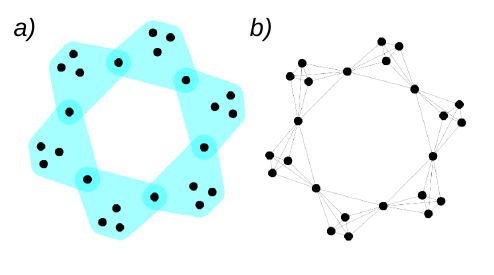}
\caption{\textbf{Non-local $4$-hyperring vs. clique-projected network: topology} a) A non-local $4$-hyperring with $m=6$ hyperedges {  and $24$ nodes}. b) Corresponding clique-projected network obtained by transforming each hyperedge into a clique.}
\label{fig:5hyperrring}
\end{figure}

%%%%%%%%%%%%% 5-body
\begin{figure}[htp!]
\centering
\includegraphics[scale=1]{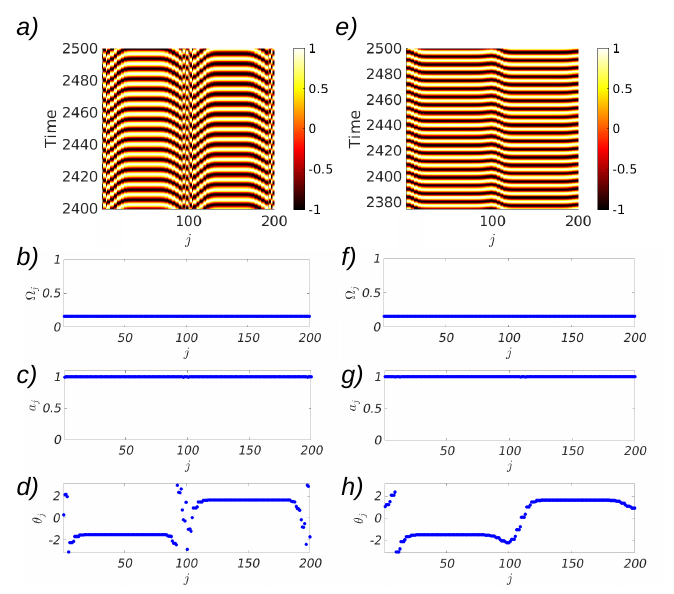}a)
\caption{\textbf{Non-local $4$-hyperring vs. clique-projected network: comparison of the dynamics} Spatiotemporal patterns for $y(t)$ and frequencies, amplitudes and phases obtained through the FFT at $2500$ time units: we observe that a chimera state with two heads is obtained in the higher-order case (panels a-d), while an almost coherent behavior is exhibited by the systems coupled through pairwise interactions (panels e-h). Let us observe that the phase varies in the interval $[-\pi,\pi)$, hence, in panel h), what we observe is not a discontinuity but it is caused by the periodicity of the phase. $j$ is the node index. {  The coupling strength is $\epsilon=0.0105$ in both cases}, the model parameters are $\lambda=1$ and $\omega=1$ and the number of oscillators is $200$.
}
\label{fig:5body}
\end{figure}

%%%%%%%%%%%%%%% 6-body + clique
\begin{figure}
\centering
\includegraphics[scale=0.88]{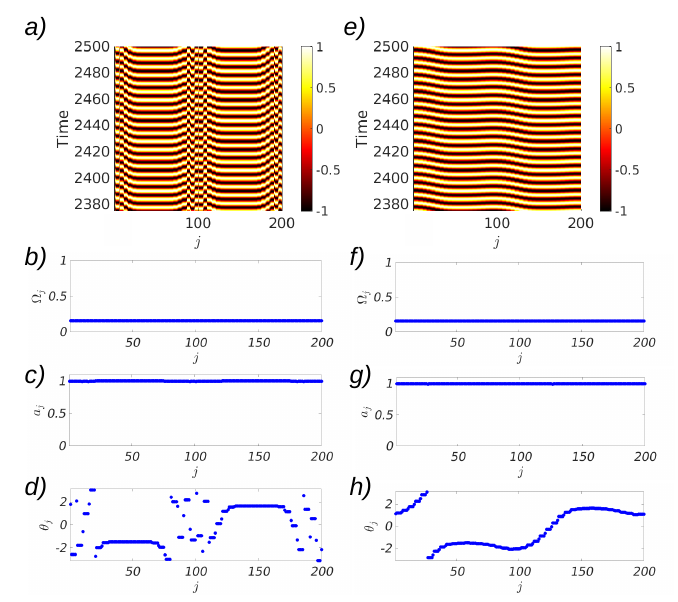}
\caption{\textbf{Non-local $5$-hyperring vs. clique-projected network: comparison of the dynamics} Spatiotemporal patterns for $y(t)$ and frequencies, amplitudes and phases obtained through the FFT at $2500$ time units: we observe that a chimera state with two heads is obtained in the higher-order case (panels a-d), while an almost coherent behavior is exhibited by the systems coupled through pairwise interactions (panels e-h). Let us observe that the phase varies in the interval $[-\pi,\pi)$, hence, in panel h), what we observe is not a discontinuity but it is caused by the periodicity of the phase. $j$ is the node index. The coupling strength is $\epsilon=0.0105$ in both cases, the model parameters are $\lambda=1$ and $\omega=1$ and the number of oscillators is $200$.}
\label{fig:6body}
\end{figure}

%%%%%%%%%%%%%%%% 3-body + clique
\begin{figure}
\centering
\includegraphics[scale=0.88]{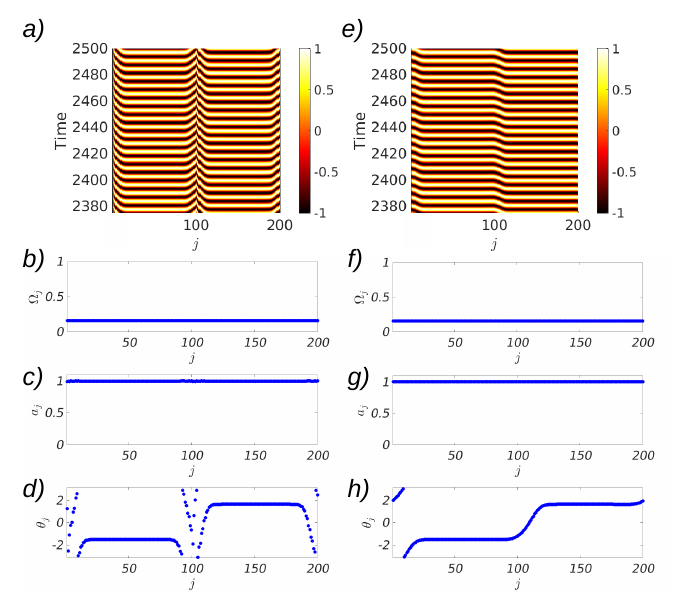}
\caption{\textbf{Non-local $2$-hyperring vs. clique-projected network: comparison of the dynamics} Spatiotemporal patterns for $y(t)$ and frequencies, amplitudes and phases obtained through the FFT at $2500$ time units: we observe that a chimera state with two heads is obtained in the higher-order case (panels a-d), while an almost coherent behavior is exhibited by the systems coupled through pairwise interactions (panels e-h). Let us observe that the phase varies in the interval $[-\pi,\pi)$, hence, in panel h), what we observe is not a discontinuity but it is caused by the periodicity of the phase. $j$ is the node index. The coupling strength is $\epsilon=0.005$ in both cases, the model parameters are $\lambda=1$ and $\omega=1$ and the number of oscillators is $200$.}
\label{fig:3body}
\end{figure}

%%%%%%%%%%%%%%%%%%%%%%%%%%%%%%%%%%%%%%%%%%%%%%%%%%%%%%%%%%%%%%%%%%%%%%%%%%%%%%%%%%%%%%%%%%%%%%%%%%%%%%%%%%%%%%%%%%%%%%%%%%%%%%%%%%%%%%%%%%%%%%%%%%%%%%%%%%%%%%%%%%%%%%%%%%%%%%%%%%%%%%%%%%%%%%%%%%%%%%%%%

%%%%%%%%%%%%%%%% necklace projections
\begin{figure}
\centering
\includegraphics[scale=1]{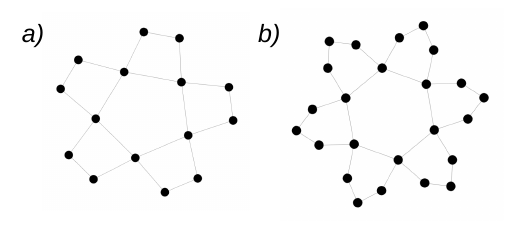}
\caption{{ \textbf{Alternative pairwise projections of non-local hyperrings} a) Projection of the non-local $3$-hyperring of Fig. 1a) (Main Text); b) projection of the non-local $4$-hyperring of Fig. \ref{fig:5hyperrring}a). Note that these are possible configurations among other ones. For instance in the case of squares, you could have joined them ``every two nodes" forming a cavity of length 8}}
\label{fig:necklace}
\end{figure}

%%%%%%%%%%%%%%%% necklace dynamics
\begin{figure}
\centering
\includegraphics[scale=0.7]{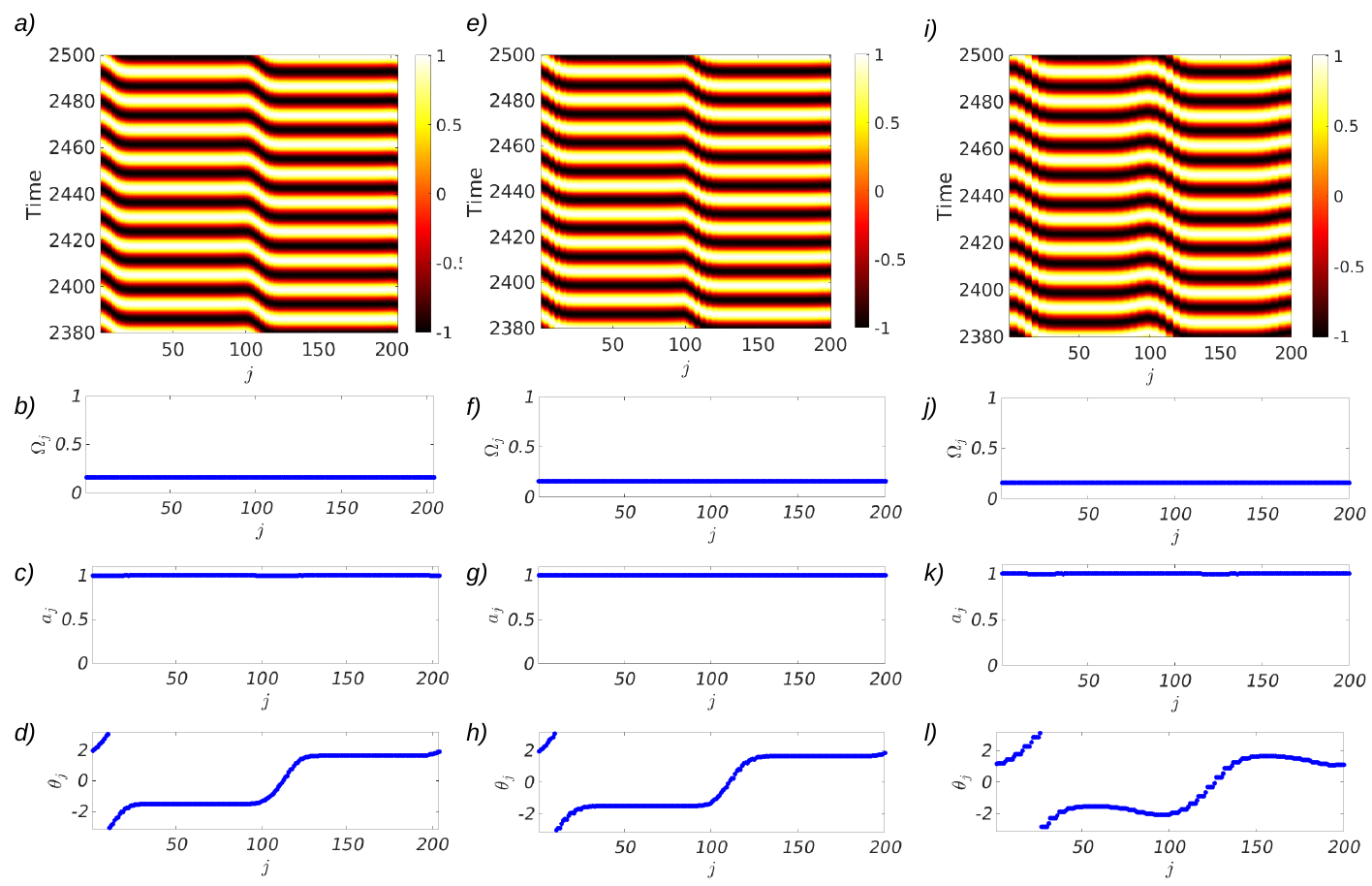}
\caption{{ \textbf{Absence of chimera states on alternative pairwise projection} Spatiotemporal patterns for $y(t)$ and frequencies, amplitudes and phases obtained through the FFT at $2500$ time units. Panels (a-d) show the case for the projection of a non-local $3$-hyperring, as in Fig. \ref{fig:necklace}a) ($204$ nodes, $\epsilon=0.01$); panels (e-h) show the case for the projection of a non-local $4$-hyperring, as in Fig. \ref{fig:necklace}b) ($200$ nodes, $\epsilon=0.0105$); panels (i-j) show the case for the projection of a non-local $4$-hyperring ($200$ nodes, $\epsilon=0.01$). We can observe that no chimera states are found in such a setting. The model parameters are $\lambda=1$ and $\omega=1$.}}
\label{fig:necklace2}
\end{figure}

%%%%%%%%%%%%%%%%% amplitude chimeras
\begin{figure}[htp!]
\centering
\includegraphics[scale=0.6]{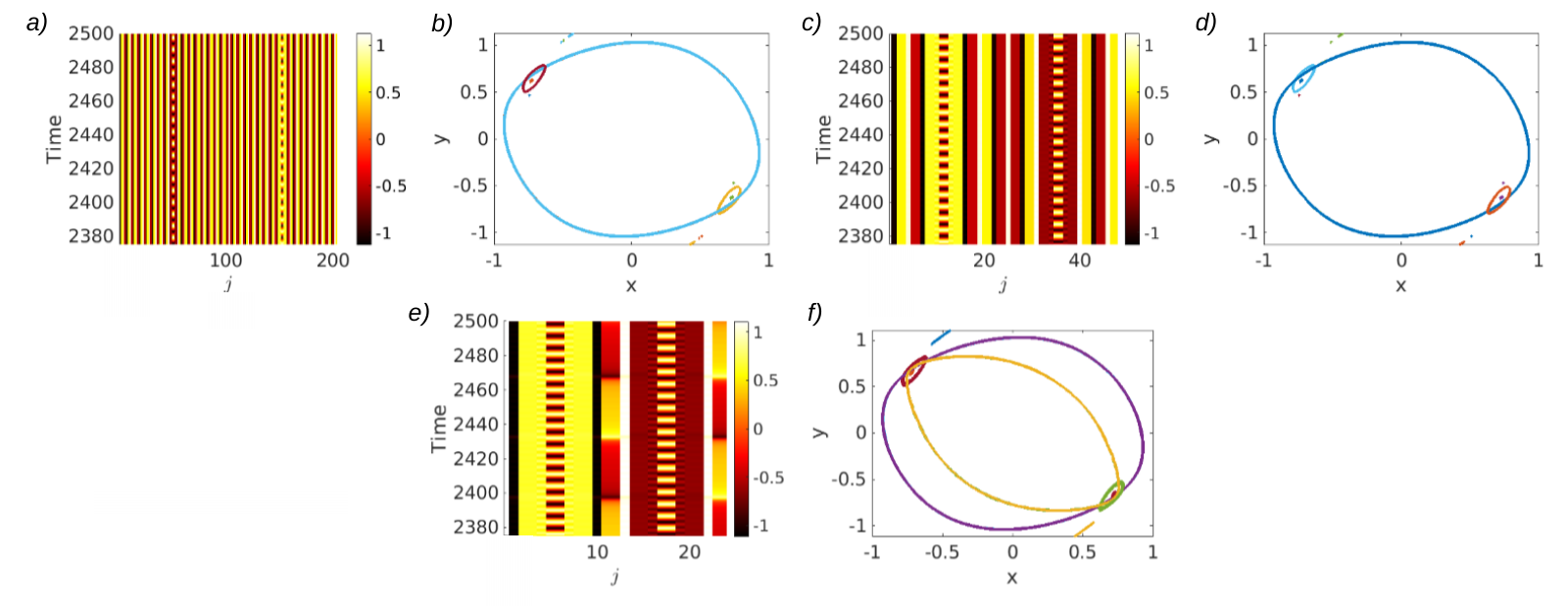}
\caption{ \textbf{Amplitude chimeras on non-local $3$-hyperrings} Panels (a,b) show the case of $204$ oscillators, panels (d,e) of $48$ oscillators, and panels (g,h) of $24$ oscillators. From the time series (panels a, c, e), and especially from the phase portraits (panels b, d, f), we can observe that some nodes are oscillating with different amplitudes, i.e., they locally exhibit an amplitude chimera patterns, while other are stationary, in a chimera death state. $j$ is the node index. The coupling strength is $\epsilon=1.1$ for all cases and the model parameters are $\lambda=1$ and $\omega=1$.}
\label{fig:ampli}
\end{figure}

%%%%%%%%%%%%%%%%%%%%%%%%%%%%%%% chimera death

\begin{figure}[htp!]
\centering
\includegraphics[scale=0.6]{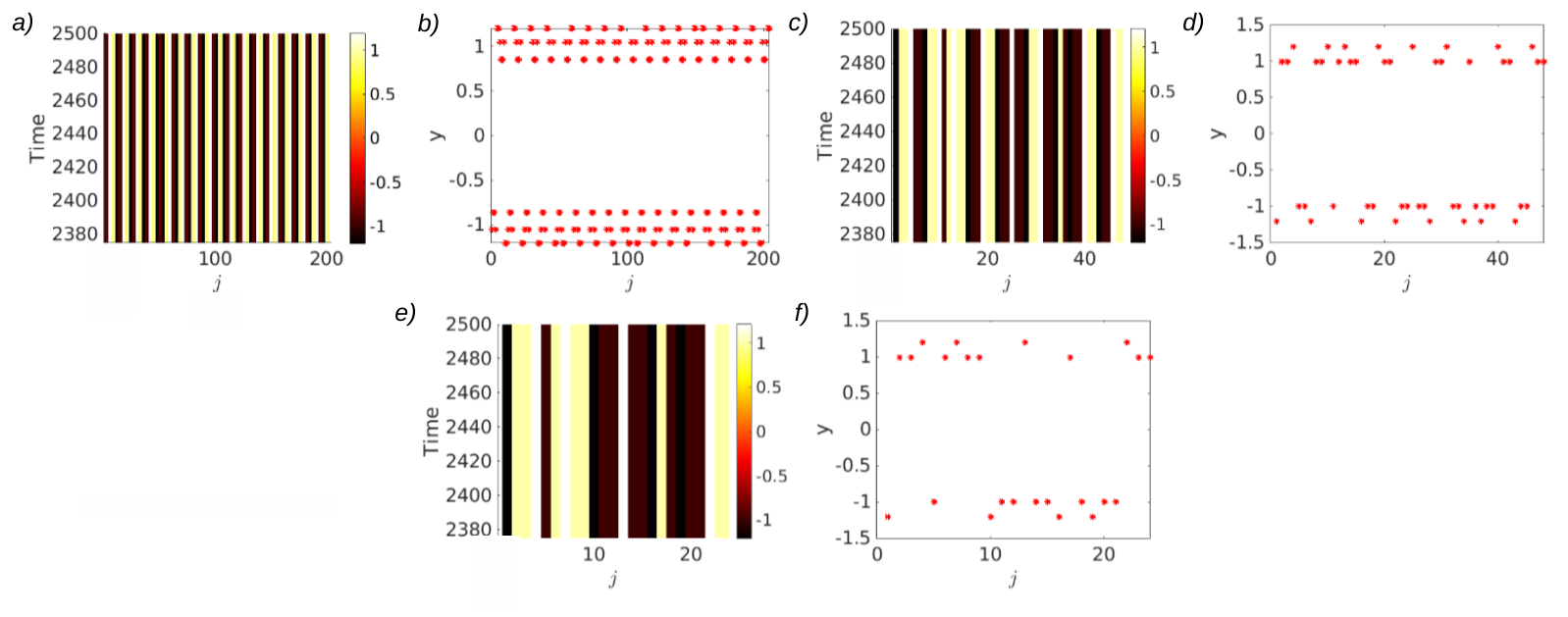}

\caption{\textbf{Chimera death on non-local $3$-hyperrings} Panels (a, b) show the case of $204$ oscillators and $\epsilon=7.7$, panels (c, d) of $48$ oscillators and $\epsilon=3.7$, and panels (e ,f) of $24$ oscillators and $\epsilon=3.7$. $j$ is the node index. The model parameters are $\lambda=1$ and $\omega=1$.}
\label{fig:death}
\end{figure}

\end{document}